\documentstyle[twocolumn,epsfig,prb,aps,amsmath,amssymb]{revtex}

\begin{document}
\draft

\twocolumn[\hsize\textwidth\columnwidth\hsize\csname
@twocolumnfalse\endcsname

\title{Low-energy excitations of spin-Peierls chains with 
modified bond-impurities}

\author{D. Augier$^a$, 
J. Riera$^{b}$ and D. Poilblanc$^a$}

\address{$^a$ Laboratoire de Physique Quantique \& Unit\'e
Mixte de Recherche 5626, Universit\'e P.~Sabatier, 31062 Toulouse,
France}

\address{$^b$ Instituto de F\'{\i}sica Rosario, Consejo Nacional de 
Investigaciones 
Cient\'{\i}ficas y T\'ecnicas y Departamento de F\'{\i}sica,
Universidad Nacional de Rosario, Avenida Pellegrini 250, 2000-Rosario,
Argentina}

\date{\today}

\maketitle

\begin{abstract}

The introduction of modified bond-defects in spin-Peierls systems
is investigated in a model of antiferromagnetic Heisenberg 
spin chains coupled to adiabatic phonons. Generically, new low-energy
magnetic or non-magnetic excitations appear below the bulk spin gap
energy. When two adjacent bonds are modified,
these excitations can be interpreted in terms of bound states 
of a soliton with the localized spin-1/2 located on the impurity site.
It is shown that the confining potential occurs even in the case of 
{\it isolated} chains. 

\end{abstract}


\vskip2pc]

\section{Introduction}

Impurity doping in quasi-one-dimensional spin-Peierls systems has
recently sparkled renewed both experimental and theoretical attention
in this field. Experimental studies in CuGeO$_3$
include so far in-chain magnetic (spin-1 Ni) or non-magnetic
(Zn, Mg) impurity doping and off-chain (Si substituting Ge) doping.
A rapid reduction of the spin-Peierls temperature has been 
generically observed as impurity concentration
increases\cite{haseimp}. Furthermore, impurity doping favors 
a new phase at low temperature, which has been characterized as a 
three-dimensional antiferromagnetic (AF) ordering by specific heat
measurements\cite{oseroff}, neutron scattering\cite{lussier} or
nuclear magnetic resonance\cite{renard}. 
More surprisingly, the AF phase
has been shown to coexist with the spin-Peierls phase for low
impurity concentrations\cite{reg,Lemmens,Weiden}.
For a theoretical understanding of these effects, the starting point 
is the realization that the elementary excitations of these 
quasi-one-dimensional systems are objects which appear on chains and
which have been characterized as topological defects
called solitons\cite{ss,Nakano,khomskii}. These solitons do not
interact with non-magnetic impurities in a strictly one-dimensional
(1D) model, but are bound to spin-1 impurities. However, impurities and
solitons are confined into bound states by the three-dimensional
character of phonons\cite{hansen}. 

Most theoretical studies have so far been concentrated on the effects
of in-chain impurities in which a Cu ion is substituted by another ion.
In this paper, we theoretically study a new class of impurity systems,
namely chains where some bonds have been modified.
This could physically correspond to the substitution of a Cu
ion by another spin-1/2 ion thus changing the values of the exchange
couplings connecting the impurity site with its two 
nearest neighbors (NN) sites. If next nearest neighbors (NNN)
couplings are included in the model, they would be modified as well.
The case of the spin-1/2 impurity could also be thought as coming
from defects in the crystal structure (structural disorder, radiation
damage, grain boundary in polycrystalline samples, etc.)
which change locally the values
of the couplings. These modified couplings are then described by an
effective spin-1/2 impurity (see e.g. Ref.~\onlinecite{Smirnov}).
Similar kind of defects could also be due to the substitution of the
out-of-chain Ge ions by Si (see e.g., Ref.~\onlinecite{reg})
then changing only one NN and two NNN bonds (``bond-centered impurity").
Since, as it was pointed out above, the elementary excitations in these
quasi-1D systems are objects which exist already in strictly 1D 
systems, the first step is to characterize these excitations by using
purely 1D models and this is the purpose of the present work. A
second step would be then to understand how these excitations and
their interactions are modified by interchain (magnetic and elastic)
interactions present in CuGeO$_3$. The first study of the effects of
bond impurities has been done on isolated Heisenberg chains in 
Ref.~\onlinecite{Eggert_PRB} using essentially field-theoretical 
techniques. We will compare our results with the ones obtained in
that study in Section~\ref{twobonds}. For spin-Peierls systems, 
analytical studies on the same model we will consider (see 
Eq.~(\ref{hamilt}) below) have been done using bosonization 
techniques\cite{Nakano,Fukuyama}. We think, however that in those
studies there is not a proper understanding of the elementary
interactions at a microscopic level.

In practice, we introduce these bond impurities in our model by
redefining the exchange coupling constants on two bonds connected
at the impurity site (spin-1/2 or ``site-centered impurity") or a
single bond (bond-centered impurity) as $J_{\rm imp}=x\, J$. 
In the first case, from a purely theoretical point of view, this
allows us to interpolate
continuously between a periodic chain with even number of sites and an
open chain with odd number of sites plus a spin disconnected to the 
chain. We find that there are also some interesting features in the 
region $x > 1$.
By assuming that lattice distortions are adiabatic,
we are thus lead to the following Hamiltonian~:
\begin{eqnarray}
{\cal H}&=&J \sum_i (1+\delta_i){\bf S}_i \cdot {\bf S}_{i+1}+ 
\alpha J \sum_i {\bf S}_i \cdot {\bf S}_{i+2}+
\frac{K}{2} \sum_l \delta_l^2
\nonumber\\
&+& J_{\rm imp}\sum_j (1+\delta_j){\bf S}_j \cdot {\bf S}_{j+1}+ 
J^{\prime}_{\rm imp} \sum_j {\bf S}_j \cdot {\bf S}_{j+2},
\label{hamilt}
\end{eqnarray}
where $i$ ($j$, $l$) indicate bulk bonds 
(impurity bonds, all bonds, respectively). We assume for simplicity
that the spring constants $K$ are not modified by the defects.
Moreover, we assume also that the two NNN couplings
at the impurity site also involve the same frustration
parameter $\alpha$, i.e. $J^{\prime}_{\rm imp}=\alpha J_{\rm imp}$. 
The variable $x$ will be extended to negative values
in which case the impurity couplings are ferromagnetic,
thus including the possibility of studying a
spin-3/2 impurity when two adjacent bonds
are modified (e.g. a Co ion replacing a Cu ion in the chain, see
Ref.~\onlinecite{Co_impur}) or a spin-1 Ni impurity if only one bond is 
changed (see Ref.~\onlinecite{Ni_impur}).
We have solved this Hamiltonian numerically on finite chains by the
self-consistent procedure described in Ref.~\onlinecite{feiguin} using
Exact Diagonalization (ED) and Quantum Monte Carlo (QMC). The relevance
of this kind of calculations to describe experimental results on
CuGeO$_3$, in particular the use of the adiabatic approximation which
somehow interpolates between the full quantum spin-lattice coupling
and the simplest models with fixed dimerization, was emphasized
in a number of previous studies.\cite{Fabrizio,Fukuyama,coupledchains}
In particular, recent calculations within this formalism\cite{sorensen}
using parameters realistic for CuGeO$_3$, have given results that 
describe Raman experiments not only at a qualitative but at a 
quantitative level, in spite of the fact that for these experiments
dynamical lattice effects are in principle important.\cite{Els}

\section{Chains with two adjacent modified bonds}
\label{twobonds}

\begin{figure}
\begin{center}
\epsfig{figure=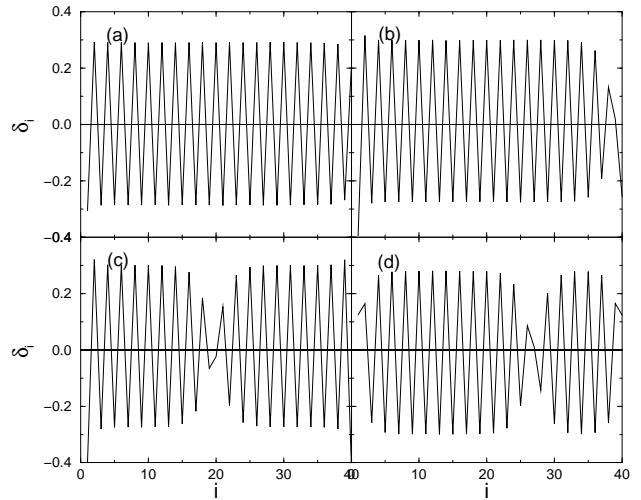,width=6.8cm,angle=-90}
\end{center}
\caption{QMC results at $T=0.05J$ 
for the distortion $\delta_i$ on a 40 site chain
with PBC, $K=2$, and for various values of $x$; (a) x=0.9, (b) x=0.3, 
(c) x=0.1, (d) x=-2.5. The spin-1/2 impurity is located at site number
40 at the extreme right of the plots.}
\label{distor12}
\end{figure}

We start our study with the case of a ``site-centered" impurity.
A frustrating exchange $J^{\prime}_{\rm imp}=\alpha J_{\rm imp}=
\alpha x J$ is also included between the impurity site and its two
next nearest neighbors. 
We first consider even chains in the region $0 \leq x \leq 1$. In the
case of $\alpha =0$, the $x=0$ point is equivalent to an open odd chain 
with a disconnected spin-1/2. 
It has been previously shown that, within this 1D model, no binding
occurs between the spin-1/2 soliton released in the chain and its
edges\cite{hansen} (which here correspond to the location of the
impurities). On the other hand, for $x=1$, we recover the pure system
and there are of course no magnetoelastic solitons in the ground state
(GS). One can then consider that the soliton is tightly bound to
the spin-1/2 impurity, forming a spin singlet. 
As $x$ is reduced from 1 to 0, this bound state is 
weakened and the soliton eventually moves away from the impurity site.
This behavior can be qualitatively seen from the QMC data at
$T=0.05J$ for $L=40$, $K=1$, as shown in Fig.~\ref{distor12}. 
In (a), $x=0.9$, the distortion pattern is almost indistinguishable
from that of a pure chain with periodic boundary conditions. In (b),
$x=0.3$, the soliton has only moved one lattice spacing away from the
impurity. In contrast, as shown in (c) corresponding to $x=0.1$, the 
bound state has disappeared and the soliton moves freely
in a region around the center of the chain segment. 

\begin{figure}
\begin{center}
\psfig{figure=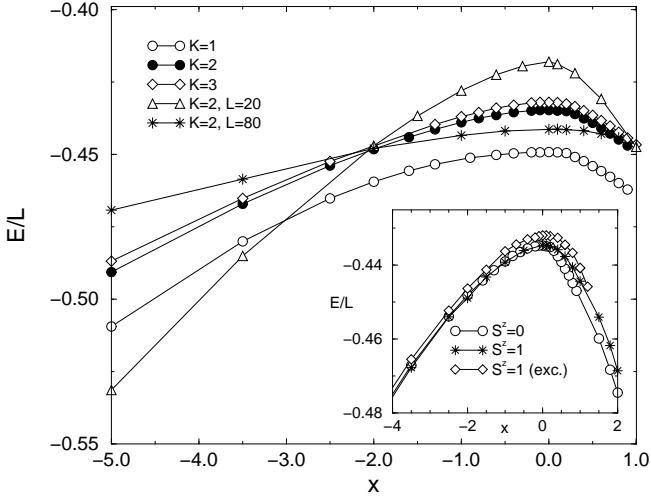,width=6.8cm,angle=-90}
\end{center}
\caption{QMC results for the energy per site as a function of $x$ on
a 40 sites
chain (except otherwise stated) with PBC and one spin1/2 impurity, 
$T=0.05$ and various values of $K$ as indicated in the figure. The 
insert shows the first two $S^z=1$ states for $K=2$ and $L=40$.}
\label{enerMCs12}
\end{figure}

To put these statements on a more quantitative basis, let us start
by analyzing the energy per site as a function of $x$, $e(x)=E(x)/L$. 
Since for $x=0$ there is no soliton-impurity binding, a
value of the energy lower than $e(x=0)$ could be considered an
indication of the presence of a bound state in the system. The QMC
results for $L=40$, $K=1, 2$ and 3, obtained for $T=0.05$ which can
be taken as the ground state energies for the system considered
here, except very close to $x=0$ as discussed below, are
shown in Fig.~\ref{enerMCs12}. For all the values of $K$
studied here it can be seen that the $e(x)$ is maximum at $x=0$.
This might indicate that there is an impurity-soliton bound state for
$x > 0$ and that the binding energy decreases as $x$ is reduced from
$x=1$, consistently with Fig.~\ref{distor12}(a), (b).
The fact that $e(x)$ is lower than $e(0$) for $x < 0$ can be
interpreted by assuming that an effective spin-3/2 impurity
starts to form and hence the
soliton is again attracted to this impurity. For the particular
value of $K=2$, we have studied the size dependence of the results
by computing the energy for $L=20$ and $L=80$ in addition to $L=40$.
The energy is always maximum at $x=0$ and, as expected, it tends to
become constant as $L$ increases. The insert shows the energy of the
lowest state in the $S^z=1$ sector together with the energy of the
excited state containing one extra soliton-antisoliton ($s-\bar{s}$)
pair. It can be seen
that for $x <0$ the ground state of the system has $S=1$. We will
discuss these features later.

The deconfinement of the impurity-soliton bound state
when $x\rightarrow 0$ ($x>0$) is a rather subtle issue.  
The above results obtained at {\it finite} temperature $T=0.05J$ and
$x=0.1$ shown in Fig.~\ref{distor12}(c) 
indicate that there is no soliton-impurity binding in that case.
However, for a temperature larger than the (zero temperature) binding
energy, one expects that, due to thermal fluctuations, the soliton can
escape from the small confining potential.
In order to confirm this scenario and to determine if the above bound
state survive in the thermodynamic limit
at zero temperature, we have computed the soliton-impurity
($T=0$) binding energy which is rigorously defined as 
$$E_B(L)= (E_{\rm imp}(L)-E_0(L)) - e_{\rm imp} - e_{\rm sol},$$
where
$e_{\rm sol}=\lim_{L \rightarrow \infty} [E_0(L+1)-E^{\ast}_0(L+1)]$,
$E_0(L)$ ($E_{\rm imp}(L)$) is the GS energy of a pure 
(bond-modified) $L$-site chain ($L=2p$ even),
$E^{\ast}_0(L+1)=(E_0(L)+E_0(L+2))/2$, and
$e_{\rm imp}=\lim_{L \rightarrow \infty}
[E_{\rm imp}(L+1)-E^{\ast}_0(L+1)]$.
These quantities have been computed within an ED 
treatment\cite{hansen}. An impurity-soliton bound state 
corresponds to $E_B(\infty)<0$. 

\begin{figure}
\begin{center}
\epsfig{figure=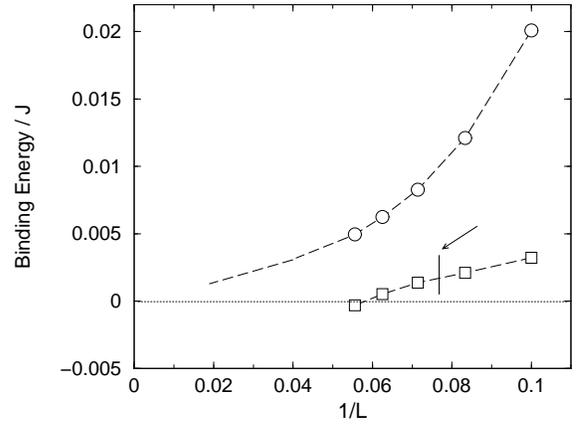,width=7cm}
\end{center}
\caption{Impurity-soliton bound state energy as a function of the
inverse lattice size $1/L$ for
$\alpha=0$, $K=J$ and x=0.0 ($\circ$) or x=0.1
($\square$)  obtained by ED. An exponential fit is also plotted
for the case $x=0$. The arrow points out the crossover between two
scaling behaviors.}
\label{ebx0}
\end{figure}

\begin{figure}
\begin{center}
\epsfig{figure=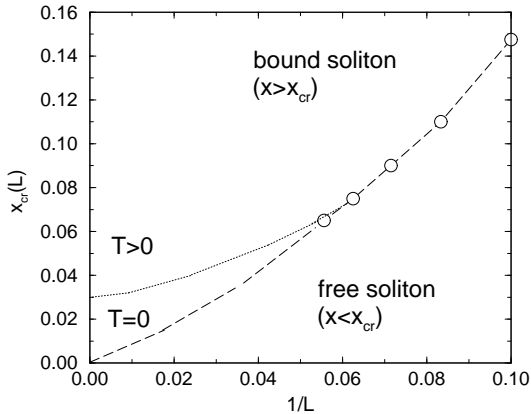,width=7cm}
\end{center}
\caption{Evaluation of critical parameter $x_{cr}$ as a function of
the inverse lattice size $1/L$ by ED. Parameters such as $\alpha=0$
and $K=J$ have been used. An
hypothetic curve at finite $T$ is also plotted.}
\label{crit}
\end{figure}

Some results for $E_B(L)$ are shown in Fig.~\ref{ebx0}. For $x=0$, by 
using extrapolations of the form $a+b/L\,\exp(-L/L_0)$, we have
obtained a vanishing binding energy (within an
estimated error of $\sim 0.001J$) as it should be, as solitons
and non-magnetic impurities do not bind in spin-Peierls
chains including lattice relaxation\cite{hansen}. 
For $x=1$, the binding energy is the energy necessary to free one
spin which is part of a singlet, i.e. the spin gap. We obtained
$E_B=-0.39J=-\Delta_{\rm spin}$, which implies that there is no 
$s-\bar{s}$ binding in this model\cite{augier}.
For $x=0.1$, the situation, as depicted also in Fig.~\ref{ebx0} is
more intricate. In this case, there is a cross-over
between the free and bound regimes near $L=14$ as signaled 
by a change of the curvature of the scaling form of the
binding energy. The exponential fit is of course no more valid but 
one can crudely estimate that in the bulk limit $E_B < -0.01J$. 
This result suggests that, for fixed positive $x$, binding sets up only
for sufficiently large chain lengths.  Alternatively, this implies
the existence of a finite critical value of $x$, 
$x_{cr}(L)$, below which there is no impurity-soliton bound state
for a given chain size $L$. 
This critical value $x_{cr}(L)$ has been evaluated by 
ED for various lattice sizes and results are shown in Fig.~\ref{crit}.
Note that this crossover disappears for
$x > 0.3$ i.e. the soliton is bound for all sizes we considered
and the extrapolated binding energy is clearly non zero (negative).
The extrapolation of the $x_{cr}(L)$ curve 
when $L\rightarrow \infty$ is consistent with a finite 
binding energy in the thermodynamic limit for any finite value of $x$. 

At finite temperature, a similar cross-over should also appear.
However, since for sufficiently large chains the thermal length becomes
the only relevant length scale, one expects that $x_{cr}(L)$ extrapolates,
when $L\rightarrow \infty$, to a (small) finite value consistently with 
the above QMC results of Fig.~\ref{distor12}(c) and in contrast with the 
T=0 case. The schematic behavior of $x_{cr}(L)$ at finite temperature
is shown in Fig.~\ref{crit} for comparison.

Let us now consider negative values of $x$, i.e. the case of a
ferromagnetic
bond. The limit $|x|>>1$ is simple to understand. In this case,
a spin-3/2 effective spin forms around the
impurity leaving a spin-1/2 in the rest of the even chain.
This behavior is qualitatively shown in (d) for $x=-2.5$.
Since, as we shall discuss more quantitatively later on, the soliton 
is weakly antiferromagnetically coupled to the spin-3/2 impurity, 
we expect the GS to be degenerate.
On the contrary, for $|x|<<1$, the small ferromagnetic couplings 
connected to the impurity site produce an effective 
interaction that weakly binds
the localized spin-1/2 to the spin-1/2 soliton into a triplet
bound state. 

\begin{figure}
\begin{center}
\epsfig{figure=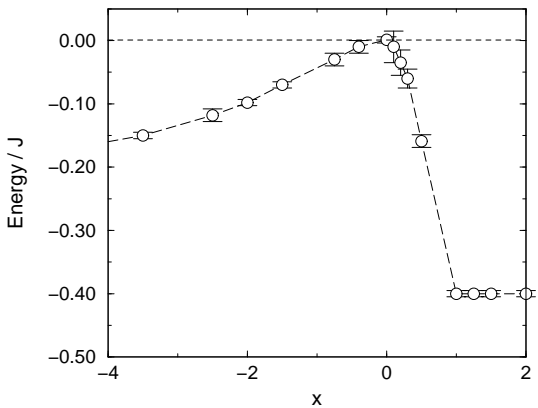,width=7cm}
\end{center}
\caption{Impurity-soliton binding energy obtained by ED
in the bulk limit as a function of x. Parameters $\alpha=0$ and 
$K=J$ have been used.}
\label{eb}
\end{figure}

In Fig.~\ref{eb} we show the zero temperature binding energy in the
thermodynamic limit as a function of $x$ together with the estimate
of the error (resulting from the finite size scaling
procedure). These results are consistent with the above
qualitative discussion and confirm the existence of a soliton-impurity 
bound state in the bulk limit, except for $x=0$.
In the $x<0$ part, a spin-3/2 is located on the impurity site and its
two neighbors, and a finite binding is also observed.
In the $x>1$ region, the soliton is very strongly bound to the
impurity, and it becomes impossible to deconfine the two objects.
In other words, trying to pull out the two components of the
singlet pair from each other will result into the spontaneous
creation of a soliton-antisoliton pair out of the ``vacuum''.
Consequently, the binding energy, as it is defined,
becomes identical to the spin gap i.e. the energy to create such 
$s-\bar{s}$ pair. 

To complete the overall picture, we have computed
the singlet-triplet spin gap defined as 
$\Delta^{01}=E(S=1,\{\delta\}_1)-E(S^z=0,\{\delta\}_0)$ where 
$\{\delta\}_0$, $\{\delta\}_1$ are the set of bond distortion
determined for $S=0$ and $S=1$ respectively. It is striking that
the overall behavior of $\Delta^{01}$ vs $x$ is totally similar to 
that of the binding energy shown in Fig.~\ref{eb}. 
This strongly suggests that around $x=0$ 
the lowest singlet ($x<0$) or triplet ($x>0$) excitation is made of
a deconfined soliton. 
When $x>1$, the spin gap is constant as the soliton remains
bound to the impurity, and bulk $s-\bar{s}$ excitations are 
energetically more favorable. 

\begin{figure}
\begin{center}
\epsfig{figure=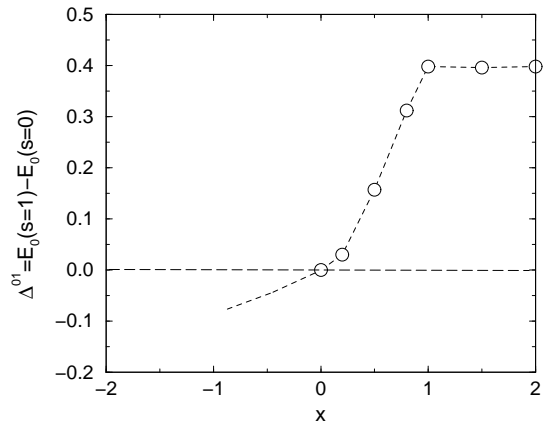,width=7cm}
\end{center}
\caption{Energy difference between the $S=1$ and
$S=0$ sectors obtained by ED in the bulk limit
as a function of $x$ for $\alpha=0$ and $K=J$.}
\label{gapk2}
\end{figure}

It is instructive to relate the above results to the analysis of
Eggert and Affleck\cite{Eggert_PRB} for a spin-1/2 Heisenberg chain
(without coupling to the lattice) in the presence of impurities.
In this work, they show that when $x > 0$, in the renormalization
group language, the open chain will be unstable and ultimately flow to
the stable periodic chain with the impurity site included (``healing"
effect). This is consistent with our result that there is a bound
state for all $x > 0$ in the bulk limit. Now, for $x < 0$, the system
would have a marginal flow (as $L \rightarrow \infty$) towards the
open chain with a decoupled spin, i.e. no impurity soliton-binding
in the bulk limit. This result is the {\it opposite} to our result
shown in Fig.~\ref{eb}. This indicates that the Luttinger liquid
approach of Ref.~\onlinecite{Eggert_PRB} cannot be directly 
extended to the case when a coupling to the lattice is present.

\section{Chains with a single modified bond}

We now turn to the case where a single bond is modified, i.e., 
on one NN bond, $J\rightarrow J_{\rm imp}$, and its two overlapping
NNN bonds, $\alpha J\rightarrow J_{\rm imp}^\prime$.
This configuration corresponds to a ``bond-centered'' rather than a 
``site-centered'' impurity. 

In the present situation, even in the case $x=0$, the number of sites
in the chain remains even and so there is no extra spin carrying
the soliton as in the previous case. Hence, QMC results for the 
distortion $\delta_i$, for $0\leq x \leq 1$, for $K=2$ and $\alpha=0$
look very similar to
those depicted in Fig.\ref{distor12}~(a). 
As it was mentioned in the Introduction, in the limit 
$x \rightarrow -\infty$, the
impurity bond leads to an effective $S=1$ impurity. On the other
hand, for $x \rightarrow \infty$, a strong dimer is formed at the
modified bond and it will be relatively decoupled from the remainder
of the chain. 
However, for intermediate values of $x$, 
it is possible that a $s-\bar{s}$ pair can be bound to the
impurity bond and, in fact, QMC results for $\delta_i$ indicate that
this may be the case for $x \leq 0$. 

\begin{figure}
\begin{center}
\psfig{figure=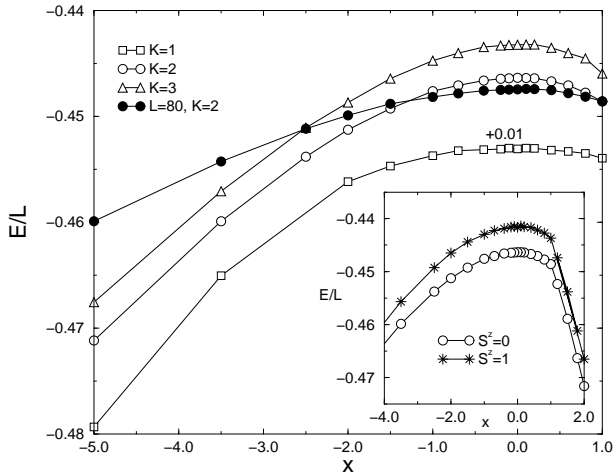,width=6.5cm,angle=-90}
\end{center}
\caption{QMC results for the energy as a function of $x$ on a 40 sites
(except otherwise stated)
chain with PBC and one modified bond, $T=0.05$ and various values of
$K$ as indicated in the figure. The insert makes a comparison of the
energies of the
lowest states in the $S^z=0$ and $S^z=1$ sectors for $L=40$, and $K=2$.}
\label{eneMCbd}
\end{figure}

If the energies of large systems
are examined (Fig.\ref{eneMCbd}), it can be seen that the highest
energy is again located at $x=0$. Notice that for $x<0$ the system
is frustrated and classically a {\em higher} energy than for $x=0$ is
expected. 
For all $x$, the GS is a singlet and the $S=1$ states are well
separated from it, as the insert of Fig.~\ref{eneMCbd} shows.
Also as this insert indicates, there is a crossover at $x=1$ between
two $S^z=0$ (in fact $S=0$) states. The ground state for 
$x > 1$ corresponds to the
above mentioned relatively decoupled dimer which leads to an energy
decreasing roughly linearly with $x$. 
It is then clear that the local ``impurity'' potential lifts the 
degeneracy of the two-fold degenerate GS occurring exactly at $x=1$
immediately as one moves away from this particular point.
Indeed, the two states crossing at $x=1$ exhibit opposite bulk 
dimerization far away from the impurity site. This can be qualitatively 
understood from the fact that, for $x<0$, a singlet is formed around
the impurity which involves three bonds weakly connected to the rest
(a triplet on the center bond bound to two solitons) compared to only
one when $x>1$. In the limit $x \rightarrow -\infty$ one recovers the
case of a spin-1 impurity (corresponding to a Ni $\rightarrow$
Cu substitution) previously studied.\cite{hansen-Ni} Precisely in 
Ref. \onlinecite{hansen-Ni} it was emphasized the role of the three-site
subsystem composed by the $\rm S=1$ site and its two $\rm S=1/2$ NN
sites.

\begin{figure}
\begin{center}
\psfig{figure=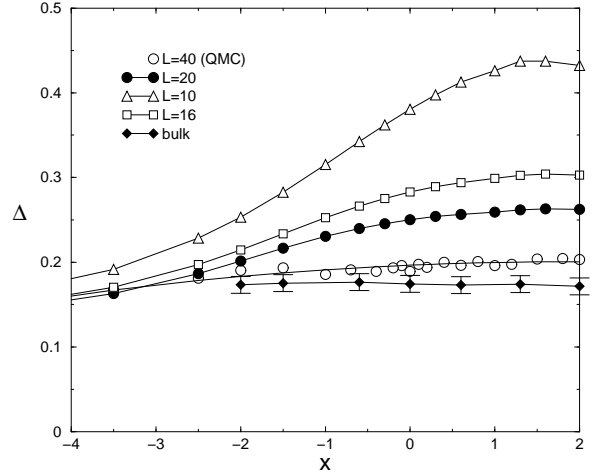,width=6.5cm,angle=-90}
\end{center}
\caption{
ED and QMC results for the spin triplet gap as a function of $x$ 
on periodic chains and one modified bond for $K=2$. The extrapolation
to the bulk is also shown in the interval $-2 \leq x \leq 2$.}
\label{gapibd}
\end{figure}

We have also computed the related spin gap for this type of impurity. 
Results obtained by both ED and QMC are shown in Fig.~\ref{gapibd}.
The extrapolation to the bulk limit

using the law $a\,\exp(-L/L_0)/L^\eta$, is also shown. It can be seen
that the gap is roughly constant (within
error bars) in the interval $-2 \leq x \leq 2$, in striking contrast
with the behavior of the site-centered impurity as discussed in the
previous Section.
It is also interesting to notice that two different triplet
excitations also cross exactly at $x=1$. Since the energy separation
from the GS is, to a good accuracy, independent of $x$ and equal to the
bulk spin gap, these states can be interpreted as a bulk (deconfined)
$s-\bar{s}$ pair excitation.

\section{Summary and conclusions}

\begin{figure}
\begin{center}
\epsfig{figure=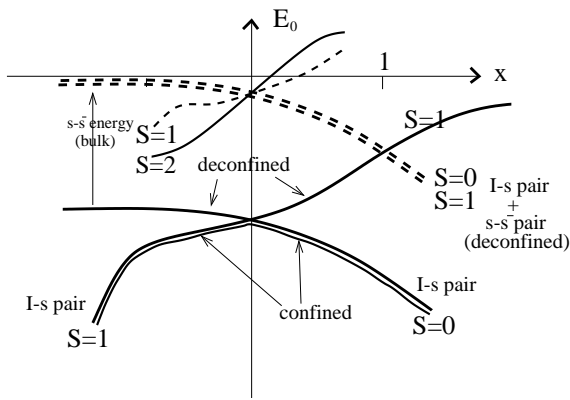,width=7.5cm}
\end{center}
\caption{Schematic representation of the low-energy spectrum of an 
even chain with a ``site-centered'' defect as
a function of $x$. The spin sectors as well as the nature of the states
are indicated in the figure.}
\label{didier}
\end{figure}

Based on the previous QMC and ED results, a schematic 
representation of the behavior of the low-energy states with $x$ 
can be drawn. A summary of the structure of the low-energy spectrum
in the case of a ``site-centered'' impurity
is schematically represented in Fig.~\ref{didier}. 
For $x>1$, the GS includes an impurity region consisting of two
adjacent strongly dimerized bonds carrying a spin-1/2 and separating
two dimerized regions with opposite $q=\pi$ lattice order parameters. 
On an even chain, this defect binds a soliton. Since, on a closed
chain, the soliton can be on either side of the defect the GS is
then two-fold degenerate as shown in Fig.~\ref{didier}.
The lowest excitation corresponds to a deconfined 
soliton-antisoliton pair decoupled from the impurity (which could be 
in singlet or triplet degenerate states) as represented by the dotted
line in Fig.~\ref{didier}.
However, for $0<x<1$, the lowest excitation is a quite different
state involving a spin flip of
the soliton spin linked to the spin-1/2 defect at the impurity site.
This qualitative change manifests itself as a level crossing of the
first excitations at $x=1$.
At exactly $x=0$, we expect additional level crossings of both GS and
excited states as shown in Fig.~\ref{didier}. Indeed, in this case, 
the first excited state 
in a chain with an even number of sites (including the impurity site)
is highly degenerate and corresponds to the 3 possible spin
configurations ($S=0,1,2$) of the three free spin-1/2 of the two mobile 
solitons and the localized spin-1/2. Arbitrary small ferromagnetic 
or antiferromagnetic couplings on the two impurity bonds are expected
to lift this degeneracy into several levels 
as seen in Fig.~\ref{didier} in agreement with the previous numerical
calculations.

\begin{figure}
\begin{center}
\epsfig{figure=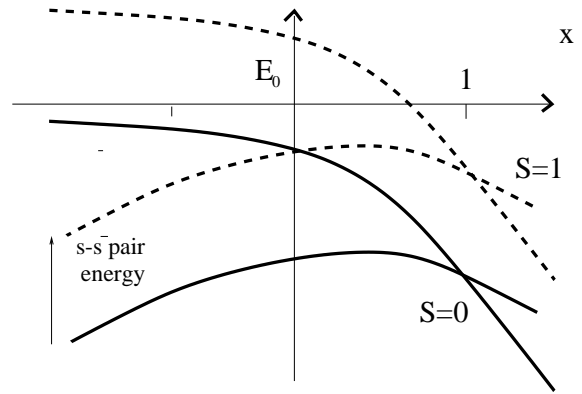,width=7.5cm}
\end{center}
\caption{Schematic representation of the low-energy spectrum of an
even chain with a ``bond-centered'' impurity as a function of $x$.
The spin sectors as well as the nature of the states
are indicated in the figure.}
\label{didier2}
\end{figure}

A similar plot of the low-energy spectrum is shown in
Fig.~\ref{didier2} in the case of a ``bond-centered'' impurity.
Contrary to the previous case, such a defect favors one of the two 
bulk dimerization patterns so that the GS is non-degenerate for all
$x$, except at $x=1$. For $x>1$ the dimerization pattern with a
singlet located on the modified (strongly dimerized) bond is selected
while, for $x<1$, strong bonds occur next to the modified one. These
two orthogonal GS naturally cross at $x=1$ as seen in
Fig.~\ref{didier2}. For all $x$, the lowest triplet excitation
corresponds to the creation of a bulk (deconfined) $s-\bar{s}$ pair.

We finish this work by discussing the implications of the above
results to experimental systems of doped spin-Peierls chains. 
Impurity doping should lead to important effects in magnetic
properties (i.e. in inelastic neutron scattering, Raman scattering,
etc...) when the spin excitation gap is
reduced, $\Delta^{01} < \Delta_{\rm spin}$ ($0<x<1$) or when the
GS carries a finite spin ($x<0$).
For defects centered on a site affecting the values of the exchange
couplings on each side, low-energy {\it magnetic} excitations are
predicted by our model as seen in Fig.~\ref{didier}.
Such states seen in inelastic neutron scattering of Zn-doped 
CuGeO$_3$ materials ($x=0$) should also survive for more general
kinds of defects such as substitution of Cu by a different spin-1/2
impurity. 

The substitution of a Ge ion by a Si one would produce a decrease
of the neighbor Cu-O-Cu angle and at the same time presumably a
shortening of the corresponding Cu-Cu distance.\cite{Weiden} 
The outcome of these competing effects on the value of the
effective NN Cu-Cu exchange
coupling is difficult to predict but it is quite likely that 
$x = 1 + \Delta_x$ with $|\Delta_x| < 1$. For both signs of 
$\Delta_x$ there is a common behavior: the impurity bond forces
the chain to pick one of the two possible dimerization patterns.
This selected pattern may or may not correspond to the one 
determined for the whole plane due to interchain magnetic
and elastic interactions. In the former case, there is a cost in 
energy
which is proportional to the length of the ``wrong" segment of
the chain (assumed finite for a finite defect density). Then,
this energy cost may be high enough to allow the formation of
a soliton in an odd chain or a $s-\bar{s}$ pair on an even
chain. The solitons will be bound to the defect and at this point
the analysis follows the one previously developed for the case
of non-magnetic impurities.\cite{coupledchains} Thus, we have for
Si-doped CuGeO$_3$ the main behavior seen for in-chain non-magnetic
impurities, namely a transition for the spin-Peierls phase to an
AF one.\cite{haseimp,oseroff,lussier,renard,reg}
Moreover, in the vicinity of $x=1$, we predict a new 
low-energy {\it singlet} excitation below the spin gap as seen in
Fig.~\ref{didier2}. Such an excitation should not be affected by
interchain interactions.
It is tempting to relate this low-energy singlet excitation to the
ones seen in Raman scattering experiments.\cite{Lemmens} However,
a more quantitative study is necessary to uniquely characterize this
excitation. Besides, other changes detected in these experiments,
like the decrease of the continuum intensity
with increasing doping were rather attributed to an enlarged NNN
interaction along the chains, an effect which was not included in 
our study.

\acknowledgments
We thank IDRIS (Orsay) for allocation of CPU time on the 
CRAY supercomputers and rs6000 workstations. J. R. also thanks
SCRI and ACNS (Florida State University) for using their computing
facilities.

\end{document}